# Zone-folded longitude acoustic phonons driving self-trapped state emission in colloidal CdSe nanoplate superlattice

Xinyu Sui[1,2], Xiaoqing Gao[3], Xianxin Wu[1,2], Chun Li[1,2,4], Xuekang Yang[2,3], Wenna Du[1,2], Zhengping Ding[5], Shengye Jin[6], Kaifeng Wu[6], Tze Chien Sum[7], Peng Gao[5], Zhiyong Tang[2,3,*], Qing Zhang[4,*], Xinfeng Liu[1,2,*]

[1]CAS Key Laboratory of Standardization and Measurement for Nanotechnology, CAS Center for Excellence in Nanoscience, National Center for Nanoscience and Technology, Beijing 100190, P. R. China

[2]University of Chinese Academy of Sciences 19 A Yuquan Rd, Shijingshan District, Beijing 100049, P. R. China

[3]CAS Key Laboratory of Nanosystem and Hierarchical Fabrication, CAS Center for Excellence in Nanoscience, National Center for Nanoscience and Technology, Beijing 100190, P. R. China

[4]Department of Materials Science and Engineering, College of Engineering, Peking University, Beijing 100871, P. R. China

[5]Electron Microscopy Laboratory, School of Physics, Peking University, Beijing 100871, China

[6]State Key Laboratory of Molecular Reaction Dynamics and Collaborative Innovation Center of Chemistry for Energy Materials (iChEM), Dalian Institute of Chemical Physics, Chinese Academy of Sciences, Dalian 116023, China

[7]School of Physical & Mathematical Sciences, Nanyang Technological University, 21 Nanyang Link, 637371, Singapore

*Email: liuxf@nanoctr.cn; q_zhang@pku.edu.cn; zytang@nanoctr.cn

**Colloidal cadmium chalcogenide nanoplates are two-dimensional semiconductors that have shown significant application potential for light-emitting technologies. The self-trapped state (STS), a special localized state originating from strong electron–phonon coupling (EPC), is likewise promising for use in one-step white light luminance owing to its broadband emission line width. However, achieving STS in cadmium chalcogenide nanocrystals is extremely challenging owing to their intrinsically weak EPC nature. By building hybrid superlattice (SL) structures via self-assembly of colloidal CdSe nanoplates (NPLs), we demonstrate in this paper zone-folded longitude acoustic phonons (ZFLAPs), which differ from monodispersed NPLs. A broadband STS emission in the spectra range of 450–600 nm is thereby observed. Through femtosecond transient absorption and impulsive vibrational spectroscopy, we reveal that STS is generated in a time scale of approximately 500 fs. It is driven by strong coupling of excitons and ZFLAPs with a Huang–Rhys parameter of approximately 22.7. Our findings provide a novel foundation for generating and manipulating STS emissions by artificially designing and building hybrid periodic structures that are superior to single-material optimization.**

Ultrathin colloidal CdX (X = S, Se, Te) nanoplates (NPLs) with a uniform atomic thickness[1-4] form a new class of two-dimensional material. They have attracted considerable attention in past decades for their excellent optical properties, such as a narrow line width[4,5] due to strong quantum thickness confinement, a high luminescent quantum yield[6] due to minimization of surface traps, and large mode gain coefficients[7-9] due to suppressed Auger recombination. These remarkable optical properties, together with their cost-effective preparation and notable stability, have made colloidal CdX NPLs promising as optoelectronic materials that can be used in light-emitting diodes[10-12] and lasers[9,13-15]. Manipulating optical transition of elementary excitations and regulating emission behaviors are fundamentally important to these NPLs.

The self-trapped state (STS) is an unusual localized state of carriers that are trapped by the deformation potential from surrounding lattices via strong electron–phonon coupling (EPC). Radiative recombination from STS has typical features, such as a large Stokes shift and a broad emission bandwidth, which engender its potential for use in one-step white-light illumination and wide gamut display[16]. STS emission has been widely observed in various condensed matter[17,18], such as rare gas, alkali halides, lead halides, and organic solids. Nevertheless, the ease of detrapping owing to thermal activation causes STS emission to occur only at low temperatures in most reported materials, which severely limits the application prospects of STS emission. Recently, several efforts to develop synthetic approaches have been made. They have focused on building a deformable lattice with considerable EPC to achieve strong room-temperature STS emission[19-23]. However, a general material design principle remains to be explored.

For colloidal CdX NPLs, the realization of room-temperature STS emission can further manifest their optoelectronic application range and promote their "lab-to-fab" transformation. Nonetheless, this objective remains a long-standing challenge. Both theories and experiments have shown that CdX nanoplates have intrinsic rigid lattices and weak EPC[24,25], which is unfavorable for the formation of STS. Increasing the fluctuation of a rigid lattice, e.g., by reducing the geometric scale of II-VI colloidal nanocrystals into an atom-cluster regime, is a possible solution to the above problem; however, it suffers from poor structural stability [26-29].

To relax restructuring of the lattice softness, which determines the STS luminescent performance, a desirable solution is to construct a hybrid SL structure constituted by self-assembly of colloidal nanocrystals. It has been variously reported in hybrid SLs of silver or gold nanocrytstals[30,31], cobalt nanocrystals[32,33], CdSe nanocrystals[34] and 2D perovskites[35], that new acoustic phonon modes called coherent longitude phonons can appear on account of the coherent motion of nanocrystals interconnected by surface ligands. It is feasible in principle to use the collective acoustic phonon modes of the hybrid SLs to be strongly coupled with excitons to generate STS.

In this paper, we therefore demonstrate a broadband STS emission in the green spectral range (519 nm, approximately 227 meV) in self-assembled colloidal CdSe NPL superlattices (SLs) that have strong EPC with a Huang–Rhys parameter of approximately 22.7. Systematic optical spectroscopy studies, including those addressing photo-excited electron paramagnetic resonance (EPR), transient absorption (TA), and low-frequency Raman spectroscopy, revealed that a self-trapped hole center is generated; this further leads to the formation of STS in the time scale of approximately 500 fs owing to the strong coupling of excitons with zone-folded longitude acoustic phonons (ZFLAPs). Unlike to the reported coherent longitude phonons, ZFLAPs are formed by the coherent superposition of acoustic phonons from CdSe NPL lattices that are separate from the coherent movement of the NPL body. The tightly connected organic ligands play the fundamental role of adding the inherent softness of the whole SL lattice, thereby promoting EPC enhancement. Our findings suggest that the hybrid structure of SLs with both organic and inorganic layers acts as a highly flexible system, unlike any other monodispersed nanostructure or bulk material. It thus provides a basis for developing artificially tunable EPCs and achieving customized STS emission.

**Results**

**Structure characterization of CdSe NPL SLs and monodispersed NPLs.** The SLs of colloidal wurtzite CdSe NPL were synthesized via a modified soft-template procedure according to Ref. 36 and was dispersed in toluene. The thickness of the NPL (regardless of surface ligands), $c_0$, was controlled to be approximately 1.4 nm (Supplementary Fig. 1) with a thickness of 3.5 monolayer, thereby guaranteeing the strong quantum confinement effect and producing sufficient structural stability. Van der Waals attraction between the surface ligand drives the orientation of the NPL assembly, leading to the formation of SLs (Fig. 1a)[3]. A transmission electron microscope (TEM) image (Fig. 1b; Supplementary Fig. 2) and a small-angle X-ray scattering spectrum (Fig. 1c) of these SLs show a distinct stacking structure with center-to-center distance $d_0$ of 2.85 nm. However, thickness $h_0$ of the monodispersed NPLs (Fig. 1d–e) is 3.9 nm and is characterized by atomic force microscopy (AFM) (Fig. 1f) according to a previous report[36]. Center-to-center distance $d_0$ in NPL SLs is approximately 1 nm shorter than $h_0$ mainly owing to ligand interdigitation (map sketch in Supplementary Fig. 3). Considering the free ligand length of approximately 1.25 nm of a single NPL, the length of ligand interdigitation (approximately 1 nm) accounts for approximately 80% of the free ligand length (Supplementary Fig. 3). It thus weakens the configurational entropy of ligands and strengthens their Van der Waals interactions[37,38], leading to a more stable SL framework.

**Absorption and emission of CdSe NPL SLs and monodispersed NPLs.** For CdSe NPL SLs, room-temperature steady-state absorption spectroscopy (black curve, Fig. 1g) exhibits two discrete peaks at 426 nm and 451 nm, which are attributed to the electron light-hole exciton transition and electron heavy-hole exciton transition[2], respectively. A long absorption tail extending to longer wavelengths is ascribed to Rayleigh scattering from the stacking of the SL structure[39]. Photoluminescence (PL) spectroscopy (green curve, Fig. 1g) shows two emission bands: a high-energy peak at 456 nm with a full width at half-maximum (FWHM) of 75 meV, and a low-energy peak at 519 nm with FWHM of 227 meV (detailed in Supplementary Fig. 4 and Supplementary Table 1). For monodispersed NPLs, the results of the absorption and emission spectroscopy under the same test condition as the NPL SLs are shown in Fig. 1h. The emission spectra of monodispersed NPLs show a single emission centered at 451 nm with FWHM of 76 meV. The absorption spectra remain almost unchanged except for the decreased scattering tail, which is caused by the absence of SL stacking. Thus, for CdSe NPL SLs, the 456 nm emission represents the excitonic emission.

To further study the broadband emission peaking at 519 nm, we performed an unbundling process to dissociate the NPL SLs into monodispersed NPLs (see the Methods section and Supplementary Fig 5) and tracked the absorption and emission spectra changes (Supplementary Figs. 6 and 7). It is worth noting that the broadband emission peak vanishes gradually during the unbundling process and finally disappears in the condition of the monodispersed NPLs (Fig. 1f; Supplementary Fig. 7). This result clearly confirms that the below-bandgap broadband emission in NPL SLs is strongly associated with the formation of the SL structure.

**TA study of CdSe NPL SLs and monodispersed NPLs.** The transient absorption (TA) spectra of NPL SLs and monodispersed NPLs were measured using the excitation of a 400 nm pump pulse (100 fs, 2.7 uJ/cm$^2$) and a time-delayed visible supercontinuum pulse. The absorption change (ΔA) with and without the pump as a function of the delay time is plotted in Figs. 2a–b. Both NPL SLs and monodispersed NPLs show negative photo-bleach (PB) peaks at 452 nm, which are marked as **X**. According to above steady-state spectra, **X** can be safely attributed to the ground-state bleach condition of the heavy-hole exciton.

In NPL SLs, another negative bleach peak located at 510 nm is marked as **Y**. Its energy is in agreement with the observed below-bandgap broadband emission. The normalized bleach recovery dynamics of **X** and **Y** in the NPL SLs and monodispersed NPLs are presented in Fig. 2c. The average life time of each decay trace is calculated based on fitting a tri-exponential decay function convoluted by the instrument response function (see Supplementary Note 1; Fig. 8; Table 2). **X** in both samples shows an average life time of approximately 0.36 ns. For **Y** in the NPL SLs, it shows a longer lifetime at approximately 3.0 ns. In the time-resolved photoluminance (TRPL) measurements, the broadband emission corresponding to bleach **Y** shows a lifetime of approximately 71 ns, which is also longer than the excitonic emission at approximately 5 ns (Supplementary Fig. 9). The lifetime measured by TA is shorter than TRPL owing to the contribution of the non-radiation transition. Based on these results, we can deem **Y** as being the bleach peak due to the trap states other than the band-edge states.

To obtain further clues about the origin of **Y**, we varied the excitation or pump fluence over two orders (0.77 µJ/cm$^2$ to 154 µJ/cm$^2$). If **Y** was the trap state originating from extrinsic permanent defects, such as surface states, vacancies, or impurities, the saturation threshold of the trapped carrier reflected in **Y** would have been considered a lower **X**. We thus plotted the bleach maximum value of **X** and **Y** as a function of the pump fluence and normalized them, as depicted in in Fig. 2d. Interestingly, we determined that **Y** follows a synchronous saturation trend compared to **X**. Both trends are well fitted by a filling model with a relative coefficient of 0.995 (see details in Supplementary Note 2). In addition, the power-dependent emission measurements for the excitons and below-bandgap broadband in the NPL SLs also show a uniform linear growth trend (Supplementary Fig. 10a). Their PL excitation spectroscopy images show a high degree of similarity (Supplementary Fig. 10b). More importantly, X-ray photoelectron spectroscopy (Supplementary Fig. 11; Supplementary Table 3) shows a balanced element distribution of Cd and Se. All of this evidence suggests that **Y** is more likely due to an intrinsic trap state rather than arising from extrinsic permanent defects. Considering that this intrinsic trap state is strongly correlated with the SL structure, we suspect it as a self-trapped state, a special electronic state in which carriers are localized by the lattice via strong EPC.

Photo-excited EPR spectroscopy, which is a powerful method to detect the existence of STS, was employed in this study[40]. This is because the unpaired electronic state due to the lattice localization would have Zeeman splitting under an externally applied magnetic field, $B_0$. The splitting energy can be written as $g_s\mu_B B_0$ ($g_s$ represents the $g$ parameter and $\mu_B$ is the Bohr magneton). It can be detected by microwave absorption spectroscopy. As shown in Fig. 2d, under excitation of the approximately 360-nm continuous wave laser, a strong EPR line in SLs is observed at 337 mT (upper panel), corresponding to $g$ = 2.004, indicating the presence of unpaired electronic states.

To eliminate the possibility that the signal was from NPL itself, EPR spectroscopy was conducted for the monodispersed NPLs, whereby a similar signal was not observed (Supplementary Fig. 12). It is worth noting that the measured $g$ parameter had a positive shift relative to the free electron value ($g$ =2.002). According to this $g$ value, the existence of a trapped hole center was indicated[41]. In metal halides or chalcogenides, the majority of the valance hole orbitals comprise anisotropic $p$ orbitals, where the unpaired orbital components induced by crystal field splitting can give rise to a positive shift in the $g$ value [42]. Thus, we believe that it is the self-trapped hole rather than the electron that forms the STS in CdSe NPL SLs.

In addition, the trapped hole in STS could be confirmed by the decay dynamics of **X** in both the CdSe NPL SLs and monodispersed NPLs (Fig. 2c). In the TA spectra, both the electron and hole contributed to the TA bleach signal, and the weight of their contribution was inversely proportional to the effective mass. In the wurtzite CdSe crystal, the effective mass of the electron and heavy hole was 0.13 $m_0$ and 0.45 $m_0$, respectively. Thus, the occupied electron states dominated the bleach dynamics in these NPLs. We assumed that, if electron was trapped in STS, a faster decay part of **X** was expected in NPL SLs compared with monodispersed NPLs due to localization by the lattice. However, we observed almost the same decay kinetics in both samples (comparison in Supplementary Table 2), indicating the absence of the electron trapping process. Consequently, apart from our conclusion from ESR, the above analysis based on TA dynamics provides further evidence of hole trapping in the STS of NPL SLs.

The formation STS dynamics was directly linked with the lattice relaxation process. In this process, strong EPC drove the carrier and lattice, which together caused a new energy minimum position. It is considered that the lattice relaxation time was similar to that of a phonon vibration circle. Consequently, we estimated the phonon frequency and energy by extracting the formation time of STS. The formation dynamics of **X** and **Y** in the TA spectra were normalized and are shown in Fig. 2d. An ultrafast onset time of approximately 500 fs for **Y**, obtained by comparing the rising edge with **X**, indicated the formation time τ of STS [43], suggesting a phonon energy with $\Omega$ = 8.1 meV or 64 cm$^{-1}$ [44]. The first-order optical phonon energy of CdSe crystal was 200 cm$^{-1}$; therefore, this phonon energy was located in the range of the acoustic phonon band of CdSe. We further explore below the source of this acoustic phonon.

To extract the transient EPC dynamics involved in STS, we closely monitored the early part of the TA dynamics. As the STS was driven by the immediate interaction of the electronic state with the lattice vibration or deformation, the light pulse (approximately 100 fs, 400 nm) excitation observed using the TA measurement was short compared to the vibrational cycle of phonons (approximately 500 fs), leading to the coherent superposition of the electron and vibrational state in the form of a vibrational wave packet. The periodic motion of the vibrational wave packet in turn modulated the probe light absorption with the frequencies of phonons. This enabled us to access not only the STS population dynamics but also the coupled lattice vibrational information[45], a method known as the impulsive vibrational spectroscopy (Supplementary Note 3). Strong oscillations were observed in the STS bleaching region between 475 and 580 nm (Fig. 3a). The time-domain oscillatory signal was superimposed on the decay trend of the STS bleach center (510 nm), representing the contribution from coupled phonons. It was resolved by subtracting the population background. The frequencies of the phonons were then determined by fast Fourier transform (FFT), as shown in the power spectrum in Fig. 3c.

In addition, the noise spectrum from the pump laser is given as a reference background for clarity (Note S4, Supplementary Fig. 13). A set of vibrational frequencies—with $\Omega_1$ of approximately 22 cm$^{-1}$, $\Omega_2$ of approximately 48 cm$^{-1}$, $\Omega_3$ of approximately 62 cm$^{-1}$, $\Omega_4$ of approximately 80 cm$^{-1}$, and $\Omega_5$ of approximately 115 cm$^{-1}$—is labeled therein. Their average value of approximately 64 cm$^{-1}$ is in accordance with the estimated value from the STS formation time. In contrast, the same measurement for the monodispersed NPLs shows a nearly indistinguishable phonon signal compared to the laser noise (Supplementary Fig. 14), indicating that the observed phonon mode was generated from the structure of SLs.

To further verify the vibrational dynamics in STS, we tuned the pump wavelength to 500 nm to achieve a near-resonant excitation with STS (Supplementary Fig. 15). Similar frequencies, such as $\Omega_1$, $\Omega_4$, and $\Omega_5$, were observed in full, except for the frequencies around 51 cm$^{-1}$, suggesting hybridization of $\Omega_2$ (approximately 48 cm$^{-1}$) and $\Omega_3$ (approximately 63 cm$^{-1}$), which arose from the resonant excitation induced wave packet potential surface shift [46].

We further performed Raman spectroscopy with 532 nm excitation to verify the presence of the observed frequencies for these SLs and the monodispersed NPLs spin-coated on the silicon oxide. They both showed the first-order longitudinal optical phonon and the surface phonon peak at approximately 199 cm$^{-1}$ and ~206 cm$^{-1}$, respectively (details in Supplementary Fig. 16 and Table. S4). For CdSe NPL SLs, a set of narrow peaks located below the optical phonon band of CdSe was formed. These modes were primarily denoted as follows: LA$_1$ of approximately 25.5 cm$^{-1}$, LA$_2$ of approximately 43.0 cm$^{-1}$, LA$_3$ of approximately 60.6 cm$^{-1}$, LA$_4$ of approximately 80.2 cm$^{-1}$, LA$_5$ of approximately 100.1 cm$^{-1}$, LA$_6$ of approximately 108.4 cm$^{-1}$, and LA$_7$ of approximately 119.4 cm$^{-1}$ (Table S5), respectively[47]. These values coincide with the modes observed in the TA measurement.

In contrast, these low-frequency modes were not observed for the monodispersed NPLs. We identified these modes as the signatures of ZFLAP of SLs. In the SL structure, the bulk longitude acoustic phonon wave vector, $k_{Bulk}$, along the SL stacking direction was folded into the SL mini-Brillouin zone (mini-BZ) by $k_{SL} = k_{Bulk}/N$, where fold number $N$ was deduced by the ratio of the SL periodicity, $d_{SL}$, to the lattice period, $a_0$[48]. Zone-folding effects transformed the propagated acoustic phonons in the bulk crystal into the equivalent optical phonons of the SL, which was Raman-detectable in the mini-BZ zone center. Owing to a sufficient ligand buckling, as previously mentioned, we considered that the acoustic phonons could propagate across the whole SL structure and finally formed the ZFLAP mode. Here, the center-to-center distance, $d_0$, of SLs together with $d_{SL}$ was 2.85 nm, the monolayer thickness of CdSe, $a_0$, was 0.4 nm[49], and fold number $N$ of the first BZ was $N = d_0/a_0 = 7$. We further calculated the ZFLAP dispersion by assuming a linear chain model (Supplementary Note 4, Supplementary Fig. 17). The simulated ZFLAP dispersion is plotted in Fig. 3e, with the yellow zone indicating the Raman-detectable mini-BZ zone center. The energies of these zone-center phonon modes were 20.8 cm$^{-1}$, 41.5 cm$^{-1}$, 61.5 cm$^{-1}$, 80.7 cm$^{-1}$, 98.4 cm$^{-1}$, 113.7 cm$^{-1}$, and 121.7 cm$^{-1}$, which are in good agreement with the experimental Raman results. Thus, we suggest that STSs were generated by the coupling of the electron-heavy hole excitons and ZFLAPs in the SL structure.

**Discussion**

In STS, carriers are localized with the short-range deformation potential created by the lattice through strong EPC. Owing to the localization, the momentum conservation is relaxed and the momentum is no longer represented by an effective quantum number. Therefore, optical transition of STS involves participation of multiple phonons with a Poisson-like distribution theorized by Huang[17], and the average phonon number change $S$ is described by the Huang–Rhys parameter. Huang–Rhys parameter $S$ is an important metric in evaluating the EPC strength. According to the experimental emission spectrum, the $S$ value can be calculated through the following equation [50]:

$$\sqrt{S} = \frac{2.36\hbar\Omega\sqrt{\coth\frac{\hbar\Omega}{2k_BT}}}{\text{FWHM}}$$

where $\hbar\Omega$ is the energy of the phonons involved, $k_B$ is the Boltzmann constant, $T$ represents the temperature, and FWHM represents the width of the emission broadening in STS. In CdSe NPL SLs, substituting the FWHM as approximately 230 meV, and the average ZFALP energy as approximately 8.1 meV, $S$ is inferred to be approximately 22.7. By comparison, the Huang–Rhys parameter of monodispersed NPLs is only 0.7 (with FWHM of approximately 75 meV), which is in accordance with the weak EPC nature of CdSe NPLs suggested in previous reports[25,51,52]. The simulated emission based on the $S$ value shows good agreement with the experimental results (Supplementary Note 5, Supplementary Fig. 18), indicating the accuracy of $S$. The Huang–Rhys parameter of 22.7 in CdSe NPL SLs is less than the value reported for $Cs_2AgInCl_6$ (38.7) [19], $Sb_2S_3$ (38.5)[53], and NaCl (42.0) [54]. Nevertheless, it is close to that reported for the CdSe nanocluster (23.0) [55], for which the STS emission is also reported.

Fig. 4a presents a schematic diagram of the STS emission in CdSe NPL SLs. The capping ligands buckle under each other via the Van der Waals interaction, thus connecting inorganic CdSe NPLs into SLs. The sufficient ligand interaction (with an 80% ligand buckling depth) enables the propagation of the acoustic phonon across the SLs, thereby driving the formation of ZFLAP. This represents the coherent superposition of acoustic phonons in each NPL. Despite ZFLAP having been widely reported in inorganic SLs, ZFLAP-induced STS has rarely been reported. It is reasonable to suggest that the organic ligands not only serve as periodic building blocks in the SL but also introduce the inherent softness of the entire SL structure, leading to the pronounced deformation potential for enhanced EPC.

To present a comprehensive mechanism for STS generation, Fig. 4b shows the potential energy of electronic states with EPC as a function of the dimensionless normal coordinate of the phonon mode. Owing to the strong coupling of photoexcited excitons with ZFLAP, the excited state (ES) would relax into STS, which has displacement $\Delta Q$ in the potential minima compared with the ground state (GS). The self-trapping energy $E_{ST}$ is contributed by multi-ZFLAP participation with value of $S\Omega_{ZFLAP}$. When we treat the potential of electronic state as harmonic, $E_{ST}$ can be written as $E_{ST}=\frac{1}{2}\Omega_{ZFLAP}\Delta Q^2$; thus, displacement $\Delta Q$ can be directly linked with Huang–Rhys parameter S in the form of $\Delta Q=\sqrt{2S}$. The displacement of STS and ES is an inverse correlation with the transition strength on account of their reduced wave package overlap. This can be explained by the observed long lifetime for STS in TA and TRPL measurements. Although STS is formed by the participation of multiple phonons, STS is not a cascade of multiple phonons;

rather, it is an overall participation of multiple phonons. Hence, self-trapping time $\tau_{ST}$ is independent of the phonon number or the Huang–Rhus parameter and is equal to $2\pi/\Omega_{ZFLAP}$. The observed $\tau_{ST}$ of approximately 500 fs is in accordance with the phonon energy extracted from the time-resolved and steady-state Raman measurements of 8.1 meV.

According to the above discussion, STS emission is directly dominated by two aspects: self-trapping energy $E_{ST}$ and displacement $\Delta Q$. The factor contributing to $E_{ST}$ and $\Delta Q$ is Huang–Rhys parameter $S$ and the energy of the participating phonon. The organic ligand in CdSe NPL SLs here unambiguously added the softness of the SL lattice, contributing to the enhancement of the Huang–Rhys parameter. ZFLAP in NPL SLs is determined by the thickness of organic and inorganic building blocks. The emission line shape evolution in CdSe NPL SLs during the unbundling process (Supplementary Fig. 7) is a rough demonstration of STS manipulation. As the SL structure gradually collapses during unbundling, the elastic inter-layer coupling of NPLs disappears, thus leading to elimination of ZFLAP modes. Meanwhile, the Huang–Rhys factor of the NPL ensemble turns from 22.7 for SLs into 0.7 for monodispersed NPLs. The continuous changing of the STS emission is reflected not only in the emission strength but also in the emission position (Supplementary Fig. 7). Mapping of the emission spectrum is shown as a function of the unbundling time in International Commission on Illumination (CIE) chromaticity coordinates in Fig. 4c. The emission color turns from green (0.22,0.64) to blue (0.17,0.09). The excitonic emission color (peak at 451 nm) and broadband emission (peak at 516 nm) occupy two of the three endpoints of the CIE chromaticity coordinates, which enable the high color rendering illumination.

In summary, we determined that the STS emission in colloidal CdSe NPL SLs is driven by the strong coupling between the excitons and the ZFLAP modes. We contend that the photoexcited holes originating from excitons are self-trapped by the surrounding lattice. We directly measured the formation time of STS via transient absorption and observed the vibrational coupling of ZFLAP in the time domain STS via impulsive vibrational spectroscopy. The Huang–Rhys parameter representing the EPC strength is calculated as approximately 22.7 based on the emission line width and phonon energy. We determined that the Huang–Rhys parameter in NPL SLs increased by as much as approximately 32 compared with monodispersed NPLs, which was strongly correlated to the ligand-induced elastic interaction of interlayer NPLs in SL.

Our results demonstrate that the optical transition in colloidal nanocrystal superstructures can be influenced not only by the reported inter-nanocrystal dipole interaction,[56-58] but also by strong EPC-induced STS generation and emission. Moreover, the hybrid colloidal SLs compared with the epitaxially grown SLs showed the former's intriguing complexity in terms of lattice and excitonic dynamics, thereby highlighting their potential in optoelectronic applications and warranting further study.

## Methods

**CdSe NPLs synthesis.** Wurtzite (WZ) CdSe nanoplates (NPLs) were synthesized by the reaction of an Se powder and Cd(OAm, OLAm)$_2$ complex in the mixture solution of octylamine (OAm) and oleylamine (OLAm), as reported in Ref. 14. To achieve Se powder dispersion, 4.5 mmol (0.355 g) of Se was mixed with 2.5 mL OAm and 2.5 mL OLAm at room temperature with vigorous stirring. To obtain a Cd(OAm, OLAm)$_2$ complex, 5 mL of OAm and 5 mL of OLAm containing 1.5 mmol (0.275 g) of CdCl$_2$·2H$_2$O were heated to 120°C for 2 h and then cooled to room temperature. Se powder dispersion was then injected into an as-prepared Cd(OAm, OLAm)$_2$ complex solution at room temperature. The final reaction mixture was moderately heated to 100°C with a 2°C/min heating rate and then maintained at 100°C for 16 h. During this period, the initial solution color changed from black to turbid yellow. After a reaction, excess ethanol containing trioctylphosphine was added, and the sample was precipitated and washed. Finally, the precipitated powder was dispersed in toluene or evaporated to dry powder.

**Unbundling Procedure of CdSe NPL.** An aliquot of WZ CdSe NPLs saturated solution (200 µL) was diluted 15-fold by toluene to afford a clear yellow dispersion. Then, a solution (1 mL) of OLAm in toluene was added to the as-prepared dilution (3 mL) to achieve the total concentration of 1% v/v in toluene. The resulting WZ CdSe NPLs solution was sonicated vigorously in a bench-top ultrasonic clean machine at room temperature for 6 h.

**Transient absorption spectroscopy (TAS).** The femto-second TAS measurements were obtained by the HELIOS commercial fs-TAS system (Ultrafast Systems). A fundamental 800 nm pulse (1 kHz, 80 fs) from a Coherent Astrella regenerative amplifier was used to pump an optical parametric amplifier (Coherent, OperA Solo) to obtain the frequency-tunable pump beam across the visible region. The pump beam was severed at 500 Hz and focused at the sample with a beam waist of approximately 300 μm. A white light continuum probe beam from 430 to 775 nm (1.6 eV to 2.9 eV) was acquired (with a beam waist of 150 μm at the sample) by focusing a small part of the fundamental 800 nm beam on a sapphire window. The magic angle was set for polarization of the pump and probe. All solution sample spectroscopy measurements were obtained with 1 mm quartz cuvettes with stirring. Finally, considering the instrument response function of this system, the system had an ultimate temporal resolution of approximately 120 fs.

**Associated content**

*Supplementary Information

Supplementary Information is available.

Supplementary Note 1-5, Supplementary Figure 1-18 and Supplementary Table 1-5 are contained in the text.

**Author information**

Corresponding Authors

*E-mail: zytang@nanoctr.cn; q_zhang@pku.edu.cn; liuxf@nanoctr.cn

**Author contributions**

Z.T., X.L, and Q.Z. designed and supervised the research; X.G synthesized and provided the sample; X.S. performed the TA measurements and data analysis; X.W, performed the steady-state low-frequency Raman spectroscopy; X.Y. performed the SAXS measurements; and Z.D. and P.G. performed the HADDF imaging. All the authors wrote and revised the manuscript.

**Additional information**

Publisher's note: *Springer Nature* remains neutral with regard to jurisdictional claims in published maps and institutional affiliations

**Acknowledgments**

X.F.L. appreciates the support of the Strategic Priority Research Program of Chinese Academy of Sciences (XDB36000000) and the Ministry of Science and Technology (2016YFA0200700 and 2017YFA0205004) and the National Natural Science Foundation of China (21673054 and 11874130). Q.Z. is grateful for funding support from the Ministry of Science and Technology (2017YFA0205700; 2017YFA0304600), the National Natural Science Foundation of China (61774003 and 61521004). This work was also supported by the National Natural Science Foundation of China (61307120, 61704038, 21805188 and 11474187).

**Competing financial interests**

The authors declare no competing interests.

**Figures and captions**

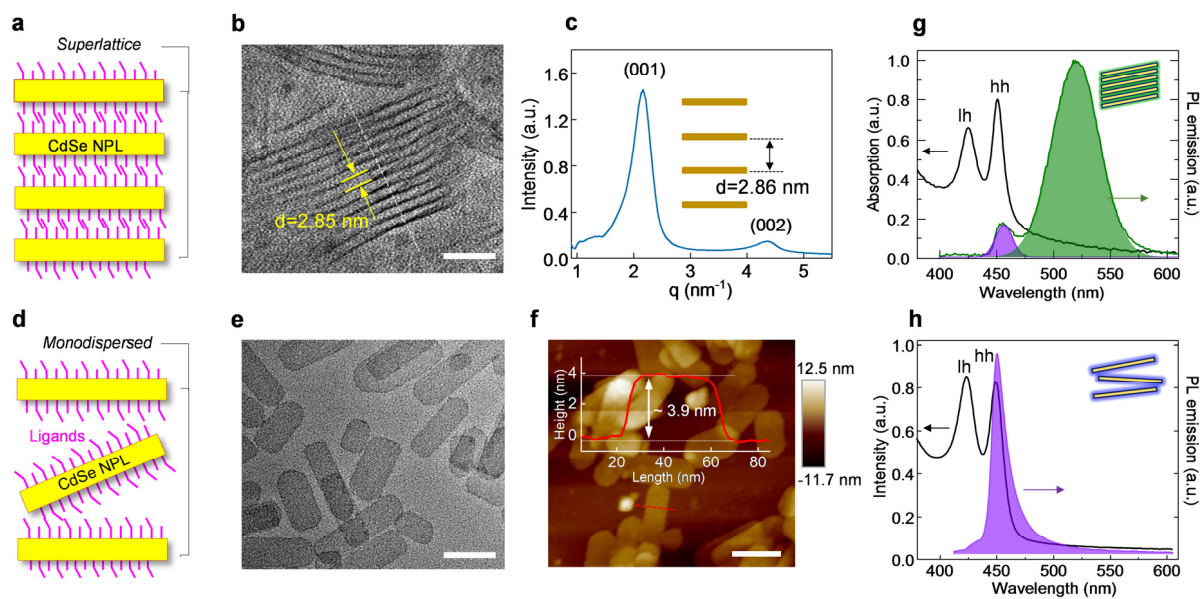

**FIG. 1. Superlattice structure-induced broadband emission.** (**a**) CdSe NPL SLs are formed from the self-assembly of CdSe NPLs through Van der Waals attraction between ligands. (**b**) Monodispersed CdSe NPLs unbundled from SLs by surfactant-assisted ultrasound method. (**c**) TEM image of CdSe NPL SLs supported by NPL edges. Yellow dashed frame indicates closely packed SLs with a center-to-center distance of 2.86 nm. (**d**) TEM image of monodispersed CdSe NPLs laid on a flat surface. Yellow dashed frame indicates a typical NPL. (**e**) SAXS pattern of colloidal CdSe NPL SLs. The first and second scattering peaks with scattering wave vector $q$ of 2.17 nm$^{-1}$ and 4.35 nm$^{-1}$ represent the contribution from (001) and (002) faces with the Bragg condition, respectively. Center-to-center distance of 2.86 nm could be calculated by the equation $L = 2\pi/q$ from reciprocal space to real space. (**f**) AFM image of monodispersed CdSe NPLs on SiO$_2$ substrate. Scale bar is 100 nm and average height is approximately 3.9 nm along the red dashed section line. (**g–h**) Room temperature absorption and PL spectra of CdSe NPL SLs and monodispersed CdSe NPLs. Both SLs and the monodispersed NPLs ensemble have an absorption peak located at 426 nm and 451 nm from the light hole and heavy hole exciton transition, respectively. Emission of SLs shows a two-peak emission located at 456 nm and 519 nm (FWHM of approximately 75 meV and 227 meV, respectively), whereas emission of monodispersed NPLs shows a single-peak emission located at 452 nm (FWHM of approximately 94 meV).

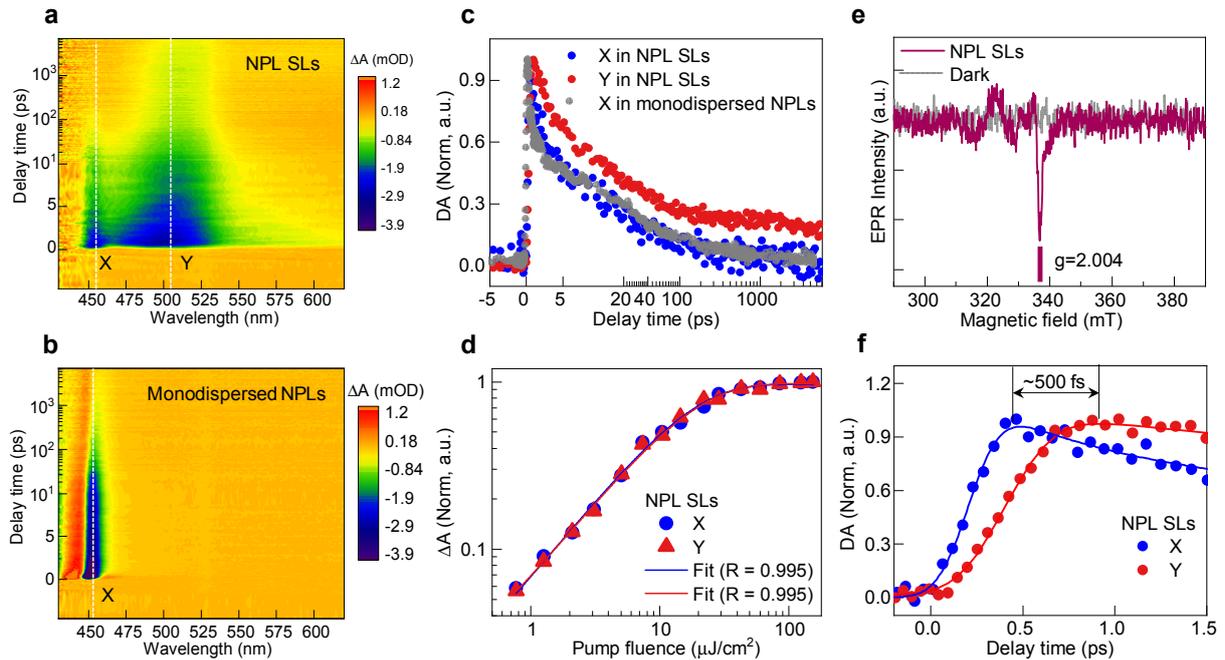

**FIG. 2. Identification of self-trapped state.** Two-dimensional contour (ΔA) plots of TA spectra of CdSe NPL SLs **(a)** and monodispersed NPLs **(b)** pumped by 400 nm with fluence of 2.8 µJ/cm$^2$. Negative peaks located 452 nm in both samples are labeled as **X** are attributed to ground state bleaching from heavy hole excitons. Negative peak located at 510 nm in CdSe NPL SLs is labeled as **Y** for further analysis. **(c)** Normalized recovery dynamics of **X** and **Y** from CdSe NPL SLs and monodispersed NPLs, respectively. **(d)** Normalized ΔA peak value of **X** and **Y** of CdSe NPL SLs as a function of pump fluence. Fitting curves based on the Poisson distribution model show a well-matched fit for relative coefficient R = 0.995. **(e)** Photoexcited EPR spectrum of CdSe NPL SLs shows a *g* parameter of 2.004 acquired by a 360-nm continuous wave laser irradiated at 77 K. **(f)** Rising delay between **X** and **Y** in CdSe NPL SLs at approximately 500 fs.

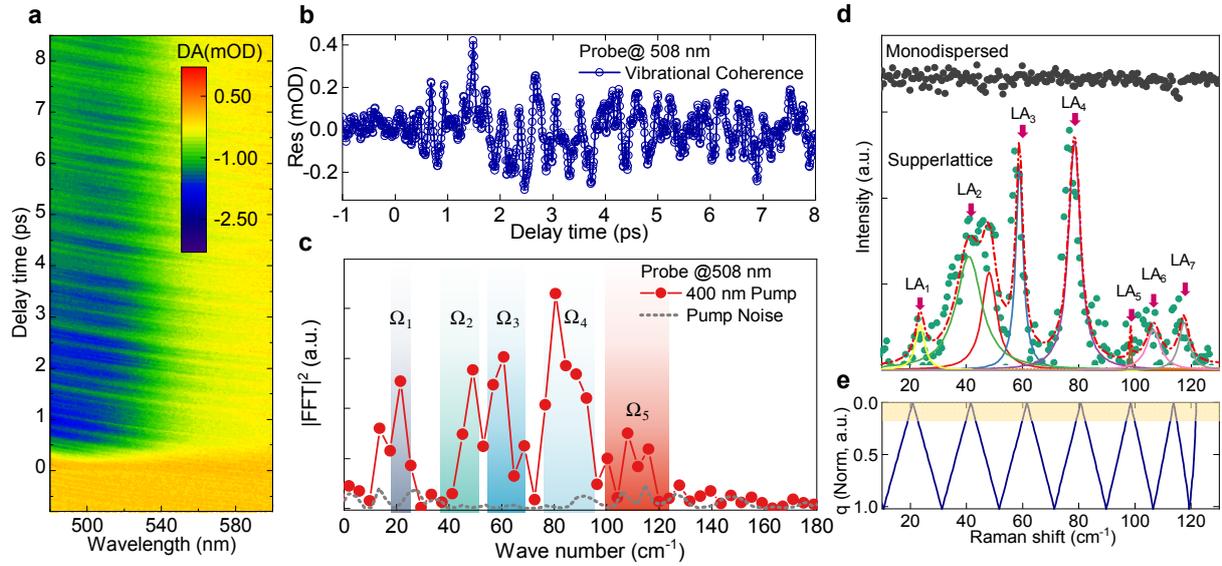

**FIG. 3. Impulsive vibration spectroscopy of self-trapped state.** (**a**) Contour plot TA spectra of CdSe NPL SLs in the first 9 ps delay window pumped at 400 nm with 4 μJ/cm². (**b**) Oscillatory components probed at 508 nm as the residual part are extracted by subtracting the population decay background. (**c**) FFT power spectrum of the above oscillatory components. A native noise background induced by laser power fluctuation is indicated by a grey dashed line as a reference. (**d**) Low-frequency Raman spectroscopy of monodispersed CdSe NPLs (grey dots) and CdSe NPL SLs (green dots). The low-frequency signals emerging from SLs are described by multi-Lorentz peak fitting. Sum of the overall fitting curve is indicated by the red dashed line. A set of arrows labeled as $LA_1$ to $LA_7$ indicates the peaks. (**e**) Calculated ZFLAP dispersion in colloidal CdSe NPL SLs based on the experimental geometry parameters. Zone center area is filled with yellow.

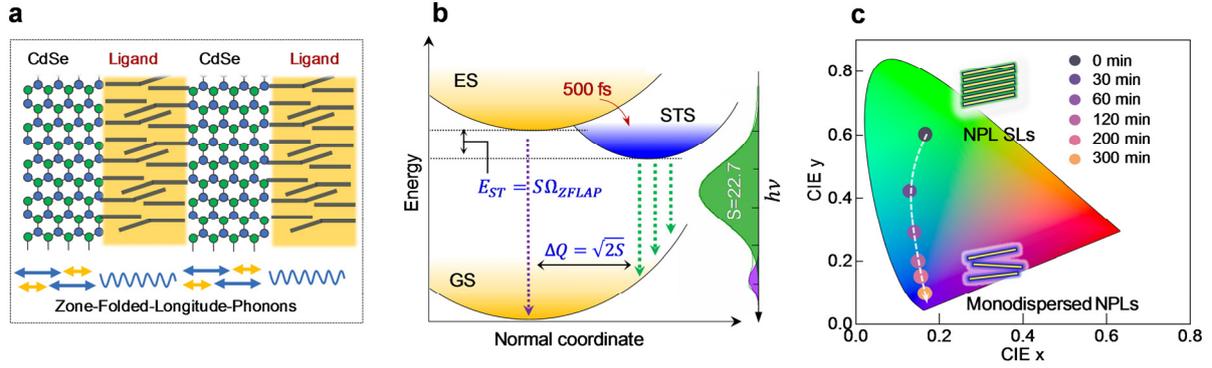

**FIG. 4. Mechanism of STS formation in CdSe NPL SLs. (a)** Schematic diagram of CdSe NPL SL composed by an inorganic CdSe NPL framework and closely buckled organic ligand layers. With the formation of SL, zone folding effects drive the formation of ZFLAP modes across the SL structure. **(b)** Normal coordinate diagram for STS driven by exciton states (ES) and strong coupling with ZFLAP mode with self-trapping energy ($E_{ST}$) of $S\Omega_{ZFLAP}$ and displacement ($\Delta Q$) of $\sqrt{2S}$. $S$ represents the Huang–Rhys parameter, meaning the average phonon number change during recombination. **(c)** Continuous emission color changing during the unbundling process of CdSe NPL SLs from stacking into the monodispersed condition labeled in CIE chromaticity coordinates.